# INTERSTELLAR POLARIZATION AND THE POSITION ANGLE ORIENTATIONS OF SEYFERT 1 GALAXIES

Short title: Polarization of Seyfert 1 Galaxies

JENNIFER L. HOFFMAN,[1,2] RYAN CHORNOCK,[1] DOUGLAS C. LEONARD,[2,3]
AND ALEXEI V. FILIPPENKO[1]

---

[1] Department of Astronomy, University of California, 601 Campbell Hall, Berkeley, CA 94720-3411, USA; jhoffman@astro.berkeley.edu; chornock@astro.berkeley.edu; alex@astro.berkeley.edu.
[2] NSF Astronomy and Astrophysics Postdoctoral Fellow.
[3] Department of Astronomy, MS 105-24, California Institute of Technology, Pasadena, CA 91125, USA; leonard@astro.caltech.edu.

ABSTRACT

We comment on recent spectropolarimetric studies that compare the observed polarization position angles (PAs) of Seyfert 1 galaxies near H$\alpha$ with the observed orientations of their radio source axes on the sky. For a Seyfert galaxy in which scattering occurs mainly in an equatorial scattering region, one expects the polarization PA to be parallel to the radio axis, while in a case in which light scatters predominantly in the polar regions, the H$\alpha$ polarization PA should be perpendicular to the radio axis. In practice, these correlations are difficult to establish because a Galactic interstellar polarization contribution can introduce a significant uncertainty into the polarization PA determination, even when the magnitude of interstellar polarization is small. We show how such uncertainties may affect the analysis of PA alignments and present spectropolarimetric observations of a probe star along the line of sight to the Seyfert 1 galaxy Mrk 871 that allow us to assess the intrinsic H$\alpha$ polarization and PA of Mrk 871. These results suggest that spectropolarimetric observations of such probe stars should form an integral part of future Seyfert galaxy polarization studies.

*Subject headings:* polarization — scattering — galaxies: active — galaxies: Seyfert — techniques: polarimetric

1. INTRODUCTION

Spectropolarimetric observations of Seyfert 1 and 2 galaxies have played an important role in the investigation of morphological properties of active galactic nuclei (AGN). The detection of polarized broad lines in Seyfert 2 galaxies led to the development of the 'unified model' for Seyfert galaxies, according to which the observationally distinct Seyfert 1 and 2 galaxies are actually the same type of object viewed at different inclinations (see the review by Antonucci 1993 for a detailed history). In this picture, a central source of continuum light and broad emission lines is surrounded by an optically and geometrically thick dusty torus. Polar scattering regions lie above and below the torus along its axis, while the AGN's radio jet flows along the same axis out to great distances.

Seyfert 2 galaxies are those oriented to our line of sight such that we cannot view the emission region directly; instead, at optical wavelengths we see only the radiation that escapes along the poles of the torus and scatters toward us from electrons or dust in the polar scattering regions. The polarization produced by this geometry has a position angle (PA) perpendicular to the torus axis and, by extension, perpendicular to the axis of the radio source associated with the galaxy. Polarimetric observations of Seyfert 2 galaxies (e.g. Antonucci 1983; Brindle et al. 1990) support this geometrical model.

By contrast, Seyfert 1 galaxies are those seen at inclinations closer to the axis of the dusty torus. In most of these cases, the optical radiation we observe is polarized along the axis of the torus, so that the observed polarization PAs of Seyfert 1 galaxies are usually parallel to the axes of their radio sources (e.g. Antonucci 1983, 2001; Martel 1996). Such a PA orientation likely occurs when light scatters in a region perpendicular to the system axis. Recent studies suggest that this scattering region may be an equatorial ring inside and co-axial with the obscuring torus (Smith et al. 2002, hereafter S02; Smith et al. 2005). In this model, broad-line emission originates from a central rotating disc; electron scattering of this light in the surrounding ring, combined with wavelength-dependent dilution by direct line emission, gives rise to the PA rotations across broad H$\alpha$ line profiles and the polarization minima in the H$\alpha$ line cores observed in many of these objects (Goodrich & Miller 1994; Cohen et al. 1999; Cohen & Martel 2001; S02; Smith et al. 2005). Following Smith et al. (2004, hereafter S04), we refer to Seyfert 1 galaxies displaying these line polarization characteristics (but not necessarily parallel PA orientations) as 'equatorially scattered.'

Not all Seyfert 1 galaxies display the same polarimetric behaviour, however. Some seem to be essentially unpolarized, while others have optical polarization PAs perpendicular to their radio source axes (Antonucci 2001; S02). Still others have spectropolarimetric properties (though not necessarily PA orientations) consistent with those found in Seyfert 2 galaxies, including systematic increases in polarization toward shorter wavelengths and polarization peaks combined with constant PA across broad emission lines (Martel 1996; Antonucci 2001; S02; S04); following S04, we label these objects 'polar scattered.' By examining a sample of these unusual 'polar-scattered' objects, S02 and S04 refined the unified model with a 'two-component' scattering model that explains the diversity in optical polarization properties of Seyfert 1 galaxies as the result of a continuous orientation sequence. In this view, unpolarized Seyfert 1 galaxies are those viewed nearly pole-on to the dusty torus; 'equatorially-scattered' Seyfert 1 galaxies are those viewed at inclinations between 0° and 45°; and 'polar-scattered' Seyfert 1 galaxies are those viewed very near inclinations of 45°, such that the line of sight passes through the outer layers of the torus. Seyfert 2 galaxies have inclinations greater than 45°, so that the line of sight intersects the torus.

Optical polarization is thus a very useful observational tool for analysing Seyfert galaxies of both types. However, when interpreting polarimetric observations, one must take care to account for all sources of polarization that may contribute to the observations. Because polarization is a vector quantity consisting of both magnitude and orientation (PA), a compound polarization signal may have a magnitude and/or a PA very different from those of any of its component vectors. In particular, polarization by interstellar dust both in the Milky Way and in the host galaxy may contribute to the observed polarization of AGN.

It is currently very difficult to assess interstellar polarization (ISP) contributions from external galaxies; there are indications that the polarization behaviour of dust in at least some other galaxies may be quite different from that observed in our own (Leonard et al. 2002b; Clayton et al. 2004). However, the ISP arising from dust in the Milky Way is well studied (e.g. Serkowski, Mathewson, & Ford 1975) and can usually be assessed by observing distant 'probe stars' near the line of sight to the target of interest (see §3). Another common method of determining ISP is to average the catalogued polarimetric properties of stars near the target (Heiles 2000); since these stars can lie at a wide range of distances, however, this method is not as accurate as selecting only specific distant stars.

In some cases, features in the polarization spectrum may provide clues to the characteristics of the interstellar polarization contribution. The existence of polarization changes across spectral lines indicates the presence of polarization intrinsic to the object in question, since interstellar polarization changes slowly with wavelength. It does not guarantee that the ISP contribution is negligible, as is sometimes assumed. However, given a model such as those developed by S02 and Smith et al. 2005 for the intrinsic profile of a line in the polarization or position angle spectrum, one may place constraints on the ISP contribution that gives rise to the observed profile. This method has the advantage of allowing one to estimate the *total* interstellar polarization (including any contribution arising from the host galaxy), but the disadvantage of perhaps masking unexpected and potentially informative line features.

Some authors estimate the contribution of Galactic interstellar polarization to their data using the empirical relation between reddening (in magnitudes) and maximum percentage ISP found by Serkowski et al. (1975): $p_{max} \leq 9E(B - V)$. If the estimated $p_{max}$ is small compared with the measured polarization magnitude, it is often ignored. Even when the magnitude of the ISP is small, however, its position angle may contribute significantly to the observed PA. Neglecting ISP contributions can thus lead to increased uncertainties in the derived intrinsic PAs of Seyfert galaxies.

In this paper, we first show how ISP can affect PA determinations (§2) and then estimate the uncertainties in the derived intrinsic PAs of the objects in the Seyfert 1 sample of S04 (§3). Finally, in §4 we present observations of an ISP probe star for the Seyfert 1 galaxy Mrk 871 and discuss the effects of ISP contamination in this specific case.

## 2. POSITION ANGLE UNCERTAINTIES DUE TO INTERSTELLAR POLARIZATION

Because polarization is a vector quantity, it is often most straightforward to depict it as the sum of Stokes vectors in the $q$–$u$ plane; one must only keep in mind that in this plane the polarization PA between two vectors is defined as half the geometrical angle between them (Chandrasekhar 1946). Let us suppose that a given polarimetric measurement is the sum of only two components: the polarization intrinsic to the object of interest, and the polarization due to Galactic interstellar dust. (We will return to the topic of polarization by host-galaxy dust in §3.) Figure 1 shows a schematic depiction of this scenario. If the magnitude of the ISP ($p_{ISP}$) is known or estimated but its PA is unknown, then the set of possible ISP vectors can be represented by a circle with radius $p_{ISP}$. Corresponding intrinsic polarization vectors then have tails at each point on the circle and heads at the observed ($q$, $u$) point. Except when the ISP vector is parallel or antiparallel to the observed polarization vector, the intrinsic position angles will differ from that of the observed PA.

To find the maximum PA difference for a given ISP magnitude, we choose one of the two ISP vectors for which the intrinsic polarization vector is tangent to the ISP circle, as shown in Figure 1. The maximum PA difference $\Delta\theta_{max}$ is then given by geometry:

$$\Delta\theta_{max} = 0.5 \sin^{-1}\left(\frac{p_{ISP}}{p_{obs}}\right), \qquad (1)$$

where the factor of 0.5 is due to the PA definition mentioned above. Thus, if the estimated ISP magnitude is one tenth that of the observed polarization, the maximum PA error is almost 3°, while for a ratio of one half, $\Delta\theta_{max} = 15°$. Figure 2 shows the variation of $\Delta\theta_{max}$ with the ratio of ISP to observed polarization. If the two component polarization vectors have equal magnitudes, the resulting maximum PA error is 45°. However, if the ISP has a larger magnitude than the observed polarization, $\Delta\theta_{max}$ jumps to 90°, as the PA of the intrinsic polarization can then have any value.

We note for clarity that a large value of $\Delta\theta_{max}$ does not necessarily imply that the *true* difference between observed and intrinsic PA values is large. Without an estimate of the ISP position angle, one can infer only that the difference must be between 0 and $\Delta\theta_{max}$. Therefore, in cases where the ISP position angle is unknown, $\Delta\theta_{max}$ should be seen as an additional uncertainty in the derived value of the target's PA. In the next section, we investigate the effects of such uncertainties on the analysis of Seyfert 1 galaxies.

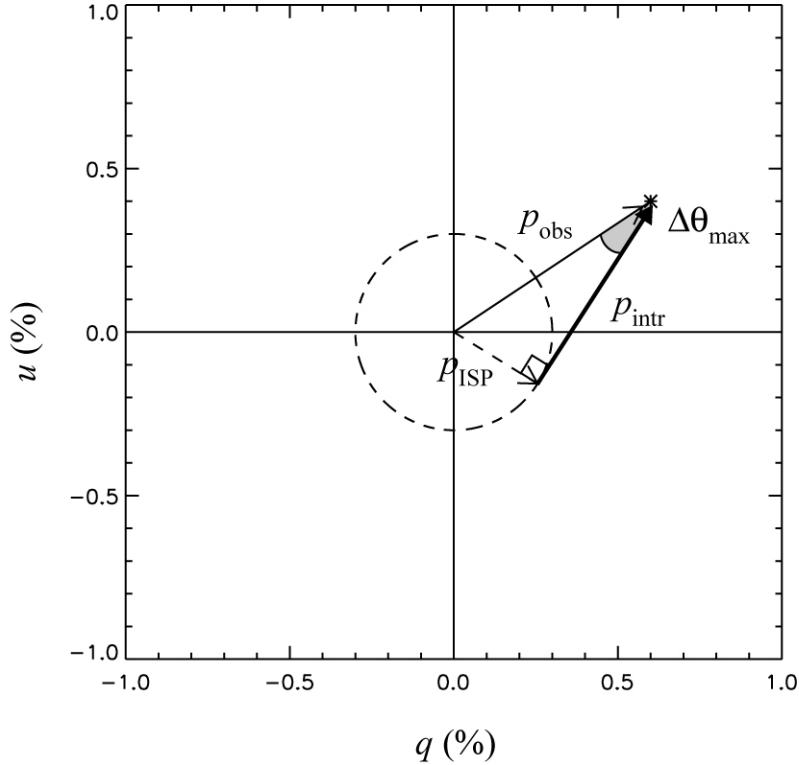

**Figure 1.** Schematic $q$–$u$ diagram showing the potential effects on the inferred PA of the target object of an interstellar polarization contribution with estimated magnitude (smaller than the observed polarization) but uncertain position angle. The solid thin arrow represents the observed total polarization. The set of possible ISP vectors is shown as a dashed circle, while one of the two ISP vectors leading to the maximum uncertainty in target PA (the shaded angle $\Delta\theta_{max}$) is shown as a dashed arrow. The intrinsic target polarization corresponding to this ISP vector is shown as a thick arrow. Another 'maximum uncertainty' ISP vector and corresponding intrinsic polarization vector exist (reflected across the line of observed polarization) but are not shown.

## 3. POSITION ANGLE UNCERTAINTIES FOR SEYFERT 1 GALAXIES

The 'two-component' scattering model developed by S02 and S04 predicts that in an 'equatorially-scattered' Seyfert 1 galaxy, the continuum polarization near H$\alpha$ should be nearly parallel to the radio source axes, while in a 'polar-scattered' object (as in a Seyfert 2 galaxy) the PA difference between the two should be ~90°. To test this prediction and investigate the relative frequencies of the two types of objects, S04 compiled from the literature and from their own observations a list of 42 Seyfert 1 galaxies with high-quality spectropolarimetric data and used this sample to investigate the agreement between the PAs of the continuum polarization near H$\alpha$ and the radio source orientations of Seyfert 1 galaxies as a group. They included only objects for which the interstellar polarization contributions (as assessed by the cited authors) appeared to be insignificant. For their own sample, Smith et al. (S02; S04) assessed the ISP contribution for each object by using its reddening (as listed in the NASA/IPAC Extragalactic

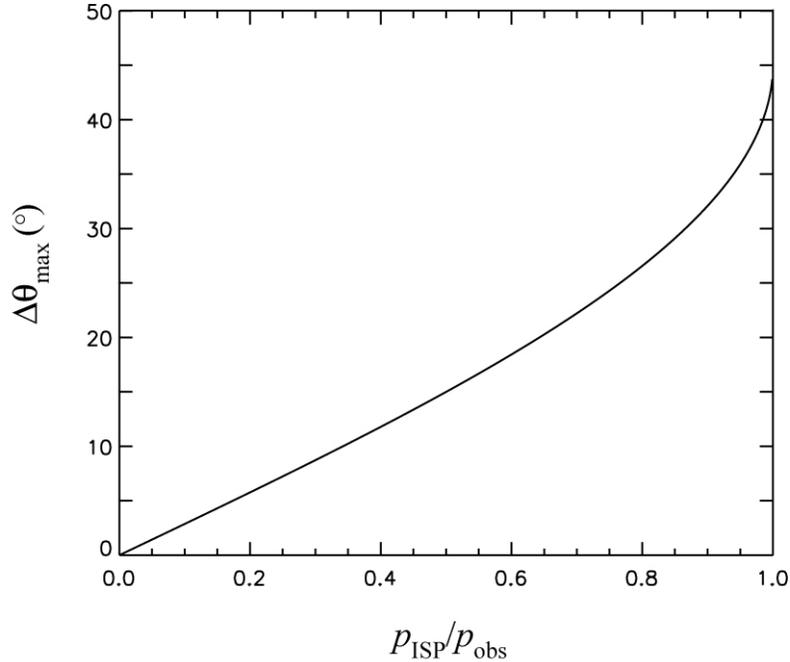

**Figure 2.** Variation of maximum PA error with the ratio of estimated ISP magnitude to observed polarization magnitude, for ratios smaller than 1.

Database [NED]) to calculate 'typical' and maximum values of $p_{ISP}$ (~ $3E(B-V)$) and ~ $9E(B-V)$, respectively) for each object.

We have calculated $p_{ISP}$ values in the same way for the other objects in the S04 sample and have applied the PA uncertainty equation from §2 to all the objects. In Table 1, we present two values of $\Delta\theta_{max}$ for each object, corresponding to these two $p_{ISP}$ estimates. We also include in Table 1 three quantities discussed in S04: the type of scattering (polar or equatorial; §1) evident in each object's polarization spectrum, the difference between the polarization and radio axis position angles ($\Delta$PA), and S04's resulting alignment classification for each object (parallel: $\Delta$PA $\leq$ 30°; intermediate: 30° $<$ $\Delta$PA $<$ 60°; perpendicular: $\Delta$PA $\geq$ 60°). We note that there is considerable uncertainty in the determination of the radio axis PAs for many of these objects (S04). We have chosen to take these PAs at face value in Table 1 so as to focus on uncertainties in polarization PA, but in the discussion below we occasionally comment on the effects of considering only those objects with unambiguous radio PAs (designated 'linear' by S04).

Six of the objects in Table 1 have been corrected for ISP by the authors who presented the original data; these are Mrk 507, Mrk 883, Mrk 957, Mrk 1126, NGC 4051, and NGC 4151. (For Mrk 883 and Mrk 1126, the authors used data from the polarization catalogues of Mathewson & Ford [1970] to make an ISP estimate; the other four were corrected using data from distant probe stars.) Of these six, three are parallel objects as defined by S04, one is perpendicular, and one is intermediate (S04 found no radio PA for Mrk 507). The only one of these objects with a well-defined polarization class is NGC 4151; it shows both an equatorial scattering signature and a parallel PA alignment, in agreement with the general unified model.

For the remaining 16 objects in Table 1 with radio PA measurements, the ratio of parallel to perpendicular to intermediate objects is 6 : 4 : 6. This ratio changes, however, if we take into account the additional PA uncertainties arising from the ISP contributions and eliminate from consideration those galaxies that can no longer be unambiguously categorized. If we assume all ISP contributions are limited to S04's 'typical' values of $3E(B-V)$, then 11 objects (those marked with single or double stars in Table 1) retain their original classifications despite the added uncertainties, for a ratio of 4 : 4 : 3; with the five ISP-corrected objects added, this ratio becomes 7 : 5 : 4. If all ISP contributions have their maximum values, the classifications necessarily remain unchanged for only four objects marked with double stars in Table 1, for a ratio of 0 : 2 : 2 (3 : 3 : 3 with the ISP-corrected objects included; 2 : 2 : 2 if we count only S04's 'linear' radio sources).

We thus find that S04's conclusion that parallel orientations outnumber perpendicular ones by 2 : 1 among Seyfert 1 galaxies, while marginally confirmed by the small sample of ISP-corrected objects, is difficult to establish with confidence for the larger sample given the PA uncertainties introduced by ISP. Measurements of ISP contributions for more of these objects will improve the reliability of this ratio and may lead to further refinement of the orientation sequence in the two-component model for Seyfert 1 galaxies.

The expected correspondences predicted by S04 between polar-scattering signatures and perpendicular orientations, and between equatorial-scattering signatures and parallel orientations, appear to hold even when we take the ISP uncertainties into account, though the reduced sample size makes statistical conclusions impractical. Of the 11 objects in Table 1 assigned both a polarization class and an alignment, 7 are either polar–perpendicular or equatorial–parallel, in accordance with S04's two-component model. This fraction becomes 5/7 for the objects whose alignments do not change for 'typical' ISP uncertainties, and 2/3 for those whose alignments do not change for maximal ISP uncertainties. Considering only S04's unambiguous 'linear' radio sources changes only the last of these fractions significantly, decreasing it to ½.

We also note the existence of several instances in each of which the added uncertainty due to ISP may actually strengthen the case for parallel or perpendicular classification; for example, the measured difference between polarization and radio PAs for Mrk 6 is 13.5°, while the value of $\Delta\theta_{max}$ for the 'typical' ISP contribution along this line of sight is 14°. A secure determination of the ISP contribution may lead to a smaller (i.e., more nearly parallel) value of ΔPA for this object. (However, if the ISP contribution for Mrk 6 is closer to the maximum possible, its alignment class could change to either intermediate or perpendicular.) Well-determined ISP contributions may also change several intermediate classifications from Table 1 into perpendicular or parallel cases, thus potentially strengthening the evidence for the two-component model.

The most reliable way to determine both the magnitude and the position angle of the ISP toward a given target is by observing one or more probe stars along the same line of sight. Ideally, probe stars should appear close in the sky to the target object (within ~1°), lie at least ~150 pc out of the Galactic plane to sample the entire ISM, and be of type A or F to avoid intrinsic polarization (Leonard et al. 2002a; Tran 1995). We have begun a campaign at Lick Observatory to obtain polarimetric observations of probe stars for the Seyfert 1 galaxies in the samples of S02 and S04, and we present results for one object, Mrk 871, in the next section.

In cases in which the contribution of polarization from interstellar dust in the host galaxy is negligible, the above method of removing ISP suffices to establish the intrinsic polarization of the object under study. However, it is notoriously difficult to assess the contribution of $ISP_{host}$, and this may seriously affect any extragalactic polarization measurements. (One can envision the $ISP_{host}$ contribution as another arrow in Figure 1 separating $p_{ISP}$ and $p_{intr}$; without knowing the magnitude or direction of this $p_{host}$ arrow, we can draw no conclusions about the magnitude or direction of $p_{intr}$.) It may be possible to use the globular clusters in some galaxies as $ISP_{host}$ probes, as demonstrated by Clayton et al. (2004), but in general one can currently only regard the possible $ISP_{host}$ contribution as an important caveat to any detailed conclusions based on extragalactic polarimetry. We consider this all the more reason to make every attempt to quantify the contribution from Milky Way dust to extragalactic polarimetric observations.

## 4. INTRINSIC POLARIZATION OF MRK 871

To illustrate the effect that ISP contamination can have on the inferred intrinsic polarization properties of a Seyfert 1 galaxy, we present our observation of HD 145730, a probe star for the Seyfert 1 galaxy Mrk 871. We chose HD 145730 as a probe star on the basis of its A3 spectral type, proximity to Mrk 871 (0.96″), and distance from the Galactic plane (Galactic latitude $b=+40°.68$ and parallax 4.43 mas [Perryman et al. 1997] imply a height of 145 pc above the plane). We observed HD 145730 on 13 March 2005 UT, using the Kast double spectrograph (Miller & Stone 1993) with polarimeter at the Cassegrain focus of the Shane 3-m telescope at Lick Observatory. The data were reduced according to the methods detailed by Leonard et al. (2001).

To determine the polarization of HD 145730 near the Hα line, we performed error-weighted least-squares fits of quadratic expressions to the total flux and Stokes $Q$ spectra and of a linear expression to the Stokes $U$ spectrum in the range 6000–7400 Å, excluding the regions 6515–6615 Å, 6840–6950 Å, and 7150–7350 Å so as to avoid the Hα absorption line and nearby atmospheric features. This analysis yielded Stokes vectors $q_{ISP} = -0.44\% \pm 0.08\%$ and $u_{ISP} = 0.02\% \pm 0.06\%$. Observations of null standard stars with the same instrument and the fact that HD 145730 is separated from Mrk 871 by 0.96″ (a relatively large distance for a probe star) lead us to estimate the systematic errors in these Stokes parameters at ~0.1%. We thus measure the total interstellar polarization in the direction of Mrk 871 to be $0.44\% \pm 0.1\%$ at a position angle of $89° \pm 7°$.

In Figure 3, we plot this ISP estimate on a $q$–$u$ diagram along with the polarization observed by S02 for Mrk 871. Error bars on $q_{ISP}$ and $u_{ISP}$ reflect the systematic error discussed above; we have taken the uncertainty in $p_{obs}$ given by S02 to be the uncertainty in their $q_{obs}$ and $u_{obs}$ as well. Even with the large error bars, Figure 3 shows that if HD 145730 is a good probe of the Galactic ISP along the line of sight to Mrk 871, then the observed Hα continuum polarization for Mrk 871 is largely but not entirely interstellar in origin, and the intrinsic polarization of Mrk 871 has a PA differing by ~77° from that of the observed polarization. Subtracting our ISP estimate from the values observed by S02, we find the intrinsic values $q_{intr} = 0.15\% \pm 0.11\%$, $u_{intr} = 0.04\% \pm 0.11\%$, $p_{intr} = 0.15\% \pm 0.11\%$, $PA_{intr} = 7° \pm 21°$.

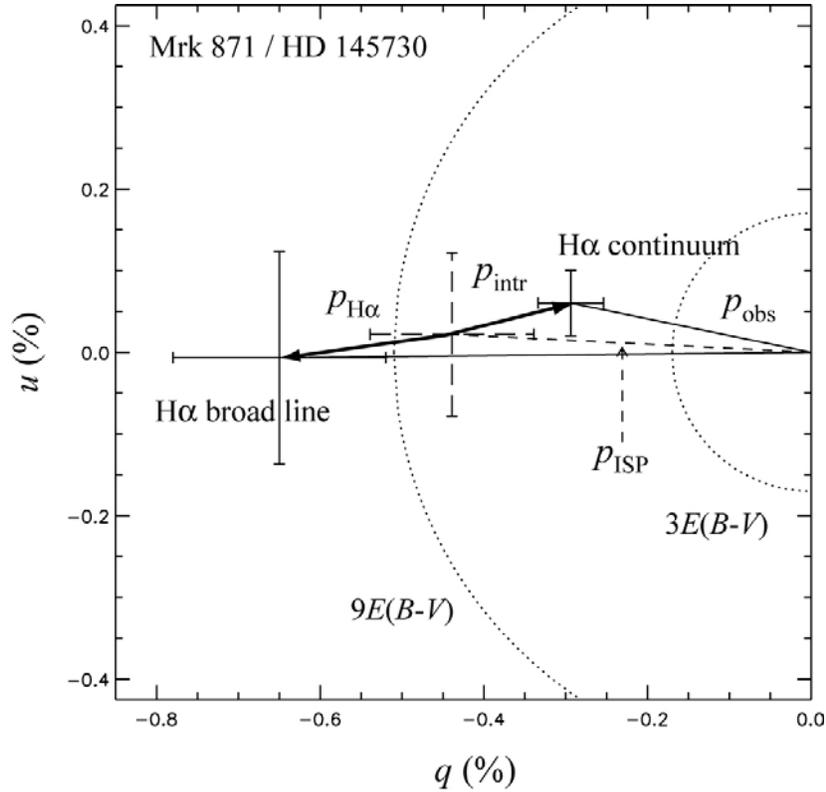

**Figure 3.** Observed and intrinsic polarization measurements for the continuum near Hα and the broad Hα emission line of Mrk 871 plotted on a *q–u* diagram. The light solid lines and points with solid error bars represent the observations of Mrk 871 published by S02; in each case, we have assumed the uncertainties on *q* and *u* to be the same as the uncertainty these authors assign to *p*. The dashed line and point with dashed error bars represent our observation of the ISP probe star HD 145730; systematic errors restrict our accuracy to ~0.1%. The heavy solid arrows indicate the resulting *intrinsic* polarization for the Hα line and nearby continuum of Mrk 871 given the ISP contribution measured for HD 145730. Dashed circles represent the sets of possible 'typical' and maximum ISP vectors for Mrk 871 one would assume based only on its reddening ($E(B-V) = 0.055$ mag; S02).

Given these results, it appears that S02 were correct to leave Mrk 871 off their list of Seyfert 1 galaxies with strong intrinsic polarization. The ISP contribution in this case is close to the maximum expected based on the measured reddening of $E(B-V) = 0.055$ mag. However, just as we caution against assuming that a small reddening value implies a negligible ISP contribution, we also note that a large reddening value—even when the ISP is correspondingly large, as in this case—does not always mean the intrinsic polarization will be small. The relative position angles of the two vectors play key roles in determining the intrinsic polarization of the object, and little can be assumed if these angles are unknown.

Finally, we point out that not only the continuum PA but also the behaviour of polarization and PA across spectral lines may be affected in some cases if ISP is not taken into account. A significant ISP contribution effectively moves the origin of polarization in the $q$–$u$ diagram; this translation can cause an apparent polarization increase or decrease across the line to be reversed once ISP is subtracted (see Leonard et al. 2000 for striking examples of different ISP estimates on the polarization spectrum of the supernova SN 1998S).

In the case of Mrk 871, we find after subtracting our ISP estimate from S02's measurements across the broad H$\alpha$ line the intrinsic polarization values $q_{H\alpha} = -0.21\% \pm 0.16\%$ , $u_{H\alpha} = -0.03\% \pm 0.16\%$, $p_{H\alpha} = 0.21\% \pm 0.16\%$, $PA_{H\alpha} = 94° \pm 22°$ (see Figure 3; these uncertainties represent the combination of the uncertainty in $p_{H\alpha}$ from S02 and the uncertainty in our ISP estimate). When ISP is taken into account, then, Mrk 871 still shows an increase in polarization across H$\alpha$, as noted by S02 using their uncorrected data. However, this increase corresponds to a difference of 44° to 88° between the position angles of the H$\alpha$ line and the continuum, which is not evident without the ISP correction. While an H$\alpha$ line polarization larger than that of the continuum is normally a feature of a 'polar-scattered' polarization spectrum, a large change in PA over the line is more consistent with 'equatorially-scattered' Seyfert 1 galaxies (S04). Further observations of Mrk 871 and nearby probe stars are warranted to constrain the ISP more closely and to investigate more fully the polarization classification of this galaxy.

## 5. CONCLUSIONS

We have shown that Galactic interstellar polarization can significantly affect position angle determinations for Seyfert galaxies, even when the magnitude of the ISP is small compared with the observed polarization, and have defined the PA uncertainty $\Delta\theta_{max}$ introduced by ISP of estimated magnitude but unknown position angle. Several objects in the sample of S04 have values of $\Delta\theta_{max}$ large enough to cause uncertainties in the assessment of the alignment of polarization PA with radio axis PA. Preliminary analysis suggests that these PA uncertainties do not adversely affect the evidence for S04's two-component model, but reliable ISP determinations could strengthen it.

Although the error bars are large, if the polarization we measure near H$\alpha$ for HD 145730 is an accurate indicator of the ISP toward Mrk 871, then this Seyfert 1 galaxy's intrinsic continuum polarization is smaller than that measured by S02 and oriented between 53° and 82° away from their estimate. We also confirm an increase in intrinsic polarization across the broad H$\alpha$ line of Mrk 871 and report a PA difference of 44° to 88° between the line and the nearby continuum.

Besides adding uncertainties to continuum PA measurements, ISP contributions may also affect apparent variations in polarization and PA with wavelength, both on large scales and across spectral lines. Since polarimetric diagnostics for Seyfert 1 galaxies often rely on such variations, one must use extreme caution in their interpretation. Our investigation argues strongly for careful consideration of ISP contributions to future polarimetric measurements of Seyfert 1 galaxies, preferably via the probe star method demonstrated here. We will present further results from our Seyfert 1 probe star observing campaign in a future contribution.

We thank M. Ganeshalingam, F. J. D. Serduke, and S. Park for obtaining the Lick observations, and J. E. Smith and an anonymous referee for providing helpful comments on drafts of this paper. JLH and DCL are supported by NSF Astronomy & Astrophysics Postdoctoral Fellowships under awards AST-0302123 and AST-0401479, respectively. AVF is grateful for a Miller Research Professorship at U.C. Berkeley, during which part of this work was completed, as well as for the support of NSF grant AST-0307894. This research has made use of the NASA/IPAC Extragalactic Database (NED), which is operated by the Jet Propulsion Laboratory, California Institute of Technology, under contract with the National Aeronautics and Space Administration, and the SIMBAD database, operated at CDS, Strasbourg, France.

TABLE 1
POSITION ANGLE UNCERTAINTIES ARISING FROM INTERSTELLAR POLARIZATION
FOR THE SAMPLE OF SMITH ET AL. (2004)

| Object | Pol. Class | ΔPA (°) | Align | $\Delta\theta_{max}$ (°) (typ ISP) | $\Delta\theta_{max}$ (°) (max ISP) | Ref | Comments |
|---|---|---|---|---|---|---|---|
| Akn 120 | EQ | 24.0 | par | 90 | 90 | 1 | ΔPA averaged over 2 observations by (2) |
| Akn 564 | … | 83.0 | perp | 10 | 90 | 1 | * |
| ESO 141-G35 | … | … | … | 18 | 90 | 1 | … |
| ESO 198-G24 | … | … | … | 4 | 11 | 3 | … |
| ESO 323-G077 | POL | … | … | 2 | 7 | 4 | … |
| Fairall 51 | POL | … | … | 2 | 7 | 1 | … |
| IRAS 15091-2107 | POL | 59.7 | int | 2 | 7 | 5 | … |
| IRAS 19580-1818 | … | … | … | 4 | 13 | 6 | … |
| IZw1 | EQ | 8.9 | par | 9 | 32 | 1 | *; ΔPA averaged over 2 observations by (2) |
| KUV 18217+6419 | EQ | 56.7 | int | 14 | 90 | 1 | … |
| Mrk 6 | EQ | 13.5 | par | 14 | 90 | 1 | * |
| Mrk 9 | … | 33.0 | int | 9 | 31 | 7 | … |
| Mrk 231 | POL | 90.0 | perp | 0 | 1 | 2 | ** |
| Mrk 279 | … | 37.1 | int | 3 | 9 | 1 | * |
| Mrk 290 | … | … | … | 2 | 5 | 1 | … |
| Mrk 304 | EQ | 86.5 | perp | 13 | 90 | 1 | * |
| Mrk 335 | … | … | … | 12 | 90 | 1 | … |
| Mrk 352 | … | … | … | 7 | 23 | 7 | $\Delta\theta_{max}$ averaged over observations at 2 telescopes |
| Mrk 376 | POL | … | … | 3 | 9 | 7 | $\Delta\theta_{max}$ averaged over observations at 2 telescopes |
| Mrk 486 | … | … | … | 1 | 2 | 7 | … |
| Mrk 507 | … | … | … | … | … | 5 | Corrected for ISP by (5) via probe star observations |
| Mrk 509 | EQ | 20.9 | par | 6 | 18 | 1 | *; ΔPA averaged over 3 observations by (2) |
| Mrk 704 | POL | … | … | 2 | 5 | 7 | $\Delta\theta_{max}$ averaged over observations at 2 telescopes |
| Mrk 766 | POL | 63.0 | perp | 1 | 2 | 5 | ** |
| Mrk 841 | EQ | … | … | 3 | 8 | 1 | … |
| Mrk 876 | EQ | 29.5 | par | 3 | 9 | 1 | … |
| Mrk 883 | … | 43.0 | int | … | … | 6 | Corrected for ISP by (6) via catalogued data |
| Mrk 957 | … | 7.3 | par | … | … | 5 | Corrected for ISP by (5) via probe star observations |
| Mrk 985 | EQ | … | … | 4 | 11 | 1 | … |
| Mrk 1048 | … | … | … | 3 | 9 | 7 | $\Delta\theta_{max}$ averaged over 2 epochs |
| Mrk 1126 | … | 72.7 | perp | … | … | 5 | Corrected for ISP by (5) via catalogued data |
| Mrk 1218 | POL | … | … | 1 | 4 | 6 | … |
| Mrk 1239 | POL | … | … | 2 | 5 | 5 | … |
| MS 1849.2-7832 | … | … | … | 8 | 29 | 1 | … |
| NGC 3227 | POL | 47.6 | int | 3 | 9 | 2 | ** |
| NGC 3516 | … | 17.0 | par | 6 | 18 | 7 | * |
| NGC 3783 | EQ | … | … | 22 | 90 | 1 | … |
| NGC 4051 | … | 12.0 | par | … | … | 7 | Corrected for ISP by (7) via probe star observations |
| NGC 4151 | EQ | 14.0 | par | … | … | 7 | Corrected for ISP by (7) via probe star observations |
| NGC 4593 | POL | … | … | 17 | 90 | 1 | … |
| NGC 5548 | … | 48.2 | int | 2 | 8 | 1 | ** |
| Was 45 | POL | … | … | 3 | 8 | 2 | … |

REFERENCES.—Original polarimetric data presented in (1) S02; (2) S04; (3) Schmid et al. (2000); (4) Schmid et al. (2003); (5) Goodrich (1989b); (6) Goodrich (1989a); (7) Martel (1996).

\* The small value of $\Delta\theta_{max}$ for this object implies that the added PA uncertainty does not affect its alignment classification for the 'typical' ISP value estimated by S02 ($3E(B - V)$). For an ISP magnitude of $9E(B - V)$, however, $\Delta\theta_{max}$ is large enough that unambiguous classification is no longer possible.

\*\* The small value of $\Delta\theta_{max}$ for this object implies that the added PA uncertainty does not affect its alignment classification for any ISP values up to $9E(B - V)$, the maximum found by Serkowski et al. (1975).

NOTE.—Column 2 describes the type of scattering (polar or equatorial; §1) implied by the object's polarization and PA spectra, from Table 4 of S04. Column 3 is the difference between polarization PA and radio axis PA from Table 4 of S04; in column 4, each source with a value in column 3 is classified as 'parallel,' 'intermediate,' or 'perpendicular' according to the definitions of S04 (see text, §3). $\Delta\theta_{max}$ values in column 4 are calculated as described in §2.